%% file: paperv05.tex
\documentclass[prd,twocolumn,superscriptaddress]{revtex4}
\usepackage{graphicx}
\usepackage{amssymb}

\input{defs}

\begin{document}
\date{\today}
\title{Deconvolution Map-Making for Cosmic Microwave
  Background Observations}
\author{Charmaine Armitage}
\email{carmitag@uiuc.edu}
\affiliation{Department of Physics, UIUC, 1110 W Green Street, Urbana, IL 61801}
\author{Benjamin D.~Wandelt}
\thanks{Benjamin D.~Wandelt is a Center for Advanced Studies Beckman Fellow}
\email{bwandelt@uiuc.edu}
\affiliation{Department of Physics, UIUC, 1110 W Green Street, Urbana, IL 61801}
\affiliation{Department of Astronomy, UIUC, 1002 W Green
Street, Urbana, IL 61801}
\affiliation{Center for Advanced Studies, UIUC, 912 W Illinois Street, Urbana,
  IL 61801}

\begin{abstract}
We describe a new map-making code for cosmic microwave background (CMB)
observations. It implements fast algorithms for convolution and transpose
convolution of two functions on the sphere (Wandelt \& G\'{o}rski 2001)
\cite{WG01}. Our code can account for arbitrary beam asymmetries and can be
applied to any scanning strategy.  We demonstrate the method using simulated
time-ordered data for three beam models and two scanning patterns, including a
coarsened version of the WMAP strategy.  We quantitatively compare our results
with a standard map-making method and demonstrate that the true sky is
recovered with high accuracy using deconvolution map-making.
\end{abstract}
\maketitle

\section{Introduction}
Real microwave telescopes collect distorted information about the cosmic
microwave background (CMB) anisotropies due to asymmetries in the beam shape
\cite{AAS02} and stray light from sources such as the Galaxy \cite{B00,B03}.
To correct for these systematic errors we must be able to remove the detector
response at all orientations of the telescope over the whole sky.  In an
optimal treatment, this correction must be applied during the map-making step
of the CMB data analysis pipeline, before the angular power spectrum can be
reconstructed.  The problem becomes increasingly important as new generations
of CMB observations probe for ever fainter signals in the CMB sky, and
especially as we are preparing to measure the polarization of the CMB with
high sensitivity. We present a complete map-making algorithm, in which
time-ordered data (TOD) is used to construct a temperature map and beam
distortions are removed.

We call our approach \textit{deconvolution map-making}, a generalization of
existing CMB map-making techniques to solve the maximum likelihood map-making
problem for arbitrary beam shapes. For sufficiently high signal-to-noise this
technique allows super-resolution imaging of the CMB from time-ordered
scans. We implement our method using the exact algorithms for the convolution
and transpose convolution of two arbitrary function on the sphere -- in this
case the sky and the beam -- as detailed by Wandelt and G\'orski in
\cite{WG01}.  These fast methods for convolution and transpose convolution are
efficient because they make use of the Fast Fourier Transform algorithm.
They are guaranteed to work to numerical precision for band-limited functions
on the sphere.

Early work on the map-making problem have relied on the brute force method of
direct matrix inversion.  However, current and future CMB experiments, like
the Wilkinson Microwave Anisotropy Probe (\textit{WMAP}) \cite{WMAP} and
Planck satellite \cite{Planck}, return enormous data sets that render the
brute force method useless.  More recent advancements include map-making
methods applicable to the latest experiments; however, many treat the beam like a
perfect delta-function (\eg \cite{D01,YM04}) or assume a symmetric beam
profile (\eg \cite{N01}), and thereby
relegate the problem of treating a non-Gaussian radial response of the
beam to subsequent stages in the data analysis \cite{mapmaking}. In this
class, special techniques exist to deal with differential measurements like
that of the Differential Microwave Radiometer (DMR) on the Cosmic Background
Explorer (\textit{COBE}) satellite \cite{lineweaver} or \textit{WMAP}
\cite{WMAPmaking}.  A Fourier method has been developed \cite{AS00} to perform
deconvolution but only for non-rotating asymmetric beams.  Lastly, \cite{BS03}
present a method to remove the main beam distortion over patches of the sky
for asymmetric, rotating beams but operate in pixel-space which is
computationally more expensive than spherical-harmonic-space algorithms for
the same level of accuracy \cite{WG01,C01}.

We test our algorithm on a simulated foreground- and Galaxy-free sky using a
standard $\Lambda$CDM power spectrum and simulated spherical harmonic
multipoles $a_{\ell m}$ up to $\ell=128$.  We also use the first-year WMAP
Ka-band temperature map as our true sky containing Galactic emission.

In section \ref{notation} we present the deconvolution method and briefly
review a standard map-making method.  In section \ref{tests} we detail the
various test cases.  Our results for the deconvolution method are given,
discussed, and compared with the standard estimates in section \ref{results}.
We conclude in section \ref{conclusions} and remark on future directions.

\section{Deconvolution Map-Making}
\label{notation}
In order to define our notation we will briefly review the path from
observations to maps.  A microwave telescope scans the CMB sky according to
some scanning strategy, effectively convolving the true sky with a beam
function, and returns a vector, $\mathbf{d}$, containing the $n_{TOD}$ samples
of the time-ordered data.  We represent this by
\begin{equation}
\mathbf{A} \mathbf{s} = \mathbf{d},
\end{equation}
where $\mathbf{A}$ is the observation matrix, defined below, and $\mathbf{s}$
is a $n_{pix}$-vector containing the true sky.

The matrix $\mathbf{A}$ encodes both the scanning strategy and the optics of
the CMB instrument.  Each sample of the TOD is modeled as the scalar product
of a row of the matrix $\mathbf{A}$ with the sky $\mathbf{s}$. Each of the
$n_{TOD}$ rows of $\mathbf{A}$ contains a rotated map of the beam. In a given
row the beam rotation corresponds to the orientation of the antenna at the
point in time when the sample is taken. We will assume the beam shape and
pointing of the satellite to be known.

The observation matrix $\mathbf{A}$ generalizes the notion of the pointing
matrix which is often used in expositions of map-making algorithms by
including both optics and scanning strategy. This generalization is necessary
for any map-making method that accounts for beam functions with azimuthal
structure.

The least-squares estimate of the true sky, $\mathbf{\hat{s}}$, is given by
\begin{equation}
\label{eq:least-squares}
\mathbf{A}^{\mathrm{T}} \mathbf{A} \mathbf{\hat{s}} = \mathbf{A}^{\mathrm{T}}
\mathbf{d}.
\end{equation}
The coefficient matrix in this system of equations, $\mathbf{A}^{\mathrm{T}}
\mathbf{A}$, is a smoothing matrix and hence ill-conditioned. Inverting it to
solve Eq.\ (\ref{eq:least-squares}) therefore poses a problem.

We describe here a regularization technique for dealing with this problem.  We
split off the ill-conditioned part of $\mathbf{A}$ by factoring the
convolution operator into $\mathbf{A} = \mathbf{BG}$ where $\mathbf{G}$ is a
simple Gaussian smoothing matrix, represented in harmonic-space by
\begin{equation}
\label{eq:gauss}
G_{\ell} = \rm{exp}\left(\frac{-\sigma^2 \ell(\ell+1)}{2}\right),
\end{equation}
where $\sigma=\mathrm{FWHM}/\sqrt{8\ln2}$.

Substituting the factorization into Eq.\ (\ref{eq:least-squares}), we get
\begin{eqnarray}
\mathbf{G}^{\mathrm{T}} \mathbf{B}^{\mathrm{T}}\mathbf{B}\mathbf{G}
\mathbf{\hat{s}}& = &\mathbf{G}^{\mathrm{T}} \mathbf{B}^{\mathrm{T}} \mathbf{d}
\label{eq:ls-gauss1}\\ \mathbf{B}^{\mathrm{T}} \mathbf{B}\mathbf{x} & = &
\mathbf{B}^{\mathrm{T}} \mathbf{d}
\label{eq:ls-gauss}
\end{eqnarray}
where we are solving for $\mathbf{x}=\mathbf{G\hat{s}}$ so as not to
reconstruct the sky at higher resolution than that of the instrument.
%The method that we will employ to solve
%for $\mathbf{x}$ (discussed below) requires a symmetric coefficient
%matrix. We have done so by removing the left-most factor $(\mathbf{G}^{\mathrm{T}})$ on  both sides of Eq.~(\ref{eq:ls-gauss}).

Equation (\ref{eq:least-squares}) is exact if the noise is stationary and
uncorrelated in the time-ordered domain.  For a more general noise covariance
matrix in the time-ordered domain, $\mathbf{N}$, the normal equation is
modified as follows
\begin{equation}
\label{eq:noise}
\mathbf{A}^{\mathrm{T}} \mathbf{N}^{-1} \mathbf{A} \mathbf{\hat{s}} =
\mathbf{A}^{\mathrm{T}} \mathbf{N}^{-1} \mathbf{d}.
\end{equation}
We proceed, considering only white noise in this paper; however, it is
straightforward to generalize to non-white noise as indicated in
Eq.~(\ref{eq:noise}). Indeed, the matrix-vector operations required for this
generalization have already been implemented in publically available
map-making codes (e.g. MADMAP \cite{madmap}).

\subsection{Fast Convolution on the Sphere}
We now briefly review the relevant formalism for general convolutions on the
sphere and refer the reader to \cite{WG01} for the full details of the fast
convolution and transpose convolution of two functions in the spherical
harmonic basis.  The convolution of a band-limited beam function
$b(\vec{\gamma})$ with the sky $s(\vec{\gamma})$ is given by the following
integral over all solid angles
\begin{equation}
\label{eq:conv-realspace}
T(\Phi_2,\Theta,\Phi_1) =
    \int{d\Omega_{\vec{\gamma}}[\hat{D}(\Phi_2,\Theta,\Phi_1)
    b](\vec{\gamma})^\ast s(\vec{\gamma})}
\end{equation}
where $\hat{D}$ is the operator of finite rotations \footnote{Our Euler angle
convention is defined as active right-hand rotations about the z, y, and z
axes by $\Phi_2,\Theta,\Phi_1$, respectively}.  In spherical harmonic space
this becomes
\begin{equation}
T_{m m' m''} = \sum_{\ell} s_{\ell m} d^{\ell}_{m m}(\theta_E) d^{\ell}_{m'
    m''}(\theta) b^{\ast}_{\ell m''}.
\label{eq:conv-harm}
\end{equation}
Analogously, the transpose convolution of $T(\Phi_2,\Theta,\Phi_1)$ is given
by
\begin{equation}
\label{eq:deconv-realspace}
%\mathrm{y}(\vec{\gamma}) = \int{d\Omega_{\vec{\gamma}}[\hat{D}(\Phi_2,\Theta,\Phi_1)
%    b](\vec{\gamma})^\ast T(\Phi_2,\Theta,\Phi_1)},
\mathrm{y}^{\ast}(\vec{\gamma}) = \int{d\Phi_2 d\Theta
    d\Phi_1[\hat{D}(\Phi_2,\Theta,\Phi_1) b](\vec{\gamma})^\ast
    T(\Phi_2,\Theta,\Phi_1)},
\end{equation} 
and in spherical harmonics
\begin{equation}
\mathrm{y}^{\ast}_{\ell m} = \sum_{m' m''} d^{\ell}_{m m'}(\theta_E)
  d^{\ell}_{m' m''}(\theta) b^{\ast}_{\ell m''} T_{m m' m''}.
\label{eq:deconv-harm}
\end{equation}

An important feature of our approach is that it economizes the computational
effort if the beam is nearly azimuthally symmetric. The parameter of the
method that sets the degree to which asymmetries of the beam are taken into
account is $m_{\rm max}$, the maximum $m''$ in equations
(\ref{eq:deconv-harm}) and (\ref{eq:conv-harm}). For $m_{\rm max}=0$ we
recover the computational cost of simple spherical harmonics transforms,
$\order{\ell_{\rm max}^3}$. Since $m_{\rm max}$ is bounded from above by
$\ell_{\rm max}$, the computational cost of the method never scales worse than
$\order{\ell_{\rm max}^4}$. For a mildy elliptical beam, we anticipate that
just including the $m_{\rm max}=0$ and $m_{\rm max}=2$ terms will suffice,
since the $m_{\rm max}=1$ term vanishes by symmetry.

For clarity, we now rewrite Eq.~(\ref{eq:ls-gauss}) in the compact
spherical-harmonic basis (summing over repeated indices)
\begin{equation}
\mathrm{A}^{\mathrm{T}}_{L' M' m m' m''} \mathrm{B}_{m m' m'' L M}
  \mathrm{x}_{L M} = \mathrm{A}^{\mathrm{T}}_{L' M' m m' m''} T_{m m' m''},
\end{equation}
where $\mathrm{A}^{\mathrm{T}}$ acting on $T_{m m' m''}$ is given by
Eq.~(\ref{eq:deconv-harm}) and $\mathrm{B}$ acting on $\mathrm{x}_{L M}$ is
given by Eq.~(\ref{eq:conv-harm}).

To make matters even more concrete, we now explicitly describe the steps
required to simulate time-ordered data $\mathbf{d}$ from a map
(``simulation''). We convolve the beam $b_{\ell m}$ with the map $a_{\ell m}$
to obtain $T_{m m' m''}$.  Then we inverse Fourier transform the $T_{m m'
m''}$ to get $T(\Phi_2,\Theta,\Phi_1)$.  Next, we must account for the scan
path $(\Phi_2(t),\Theta(t),\Phi_1(t))$, where $\Phi_2$ and $\Theta$ specify
the position on the sphere and $\Phi_1$ specifies the orientation of the beam.
This is achieved by extracting those values in $T(\Phi_2,\Theta,\Phi_1)$ which
fall on the scan path whenever the instrument samples the sky.

As a second example we describe how to compute the right hand side of
Eq.~(\ref{eq:ls-gauss}). Start with the TOD $\mathbf{d}$.  For each sample in
$\mathbf{d}$, the scanning strategy specifies the orientation
$(\Phi_2,\Theta,\Phi_1)$. The sampled temperature is added into the element of
an initially empty array which is identical in size and shape to the array
which stored $T(\Phi_2,\Theta,\Phi_1)$.  We have effectively binned the TOD
$\mathbf{d}$, according the position and orientation of the beam on the
sky. Let us therefore refer to this operation as ``binning''. In order to
minimize discreteness effects due to the gridded representation of
$T(\Phi_2,\Theta,\Phi_1)$, more sophisticated interpolation techniques could
be implemented. Additionally, the resolution of the grid into which the data
is binned may be increased.

\subsection{Solving the Deconvolution Equations}

To obtain the optimal map estimate we numerically solve the linear system of
equations in Eq.~(\ref{eq:ls-gauss}) for $\mathrm{x}_{\ell m}$. We have a choice
between direct and iterative solution methods. An iterative method is
advantageous compared to a direct method (such as Cholesky inversion) if the
cost per iteration times the number of iterations required to converge to
sufficient accuracy is less than the cost of the direct method.

For the problem sizes of current and upcoming CMB missions, where the map
contains a number of pixels $n_{pix}\sim 10^6$--$10^7$ direct solution methods
would be prohibitive for two reasons. Firstly, the required number of floating
point operations scales as $n_{pix}^3$. Secondly, the amount of space required
to store the coefficient matrix and its inverse scales as
$n_{pix}^2$. Therefore direct solution exceeds the capabilities of modern
supercomputers by several orders of magnitude. For the Planck mission direct
solution would require of order $10^{21}$ floating point operations and
hundreds of Terabytes of random access memory.

We therefore advocate using an iterative technique, the Conjugate Gradient
(CG) method \cite{CG}.  The CG method is well suited to this problem. It
solves linear systems with symmetric positive definite coefficient matrix and
has advantageous convergence properties compared to other iterative methods
such as Jacobi method \cite{CG}. In order to apply the CG method we must be
able to apply the coefficient matrix on the left hand side of
Eq.~(\ref{eq:ls-gauss}) to our current guess of the solution $\mathbf{x}$. In
order to do so we simply perform the two operations of ``simulation'' and
``binning'' in succession. The fast convolution and transpose convolution
algorithms allow computing the action of the coefficient matrix on a map
without ever having to store the matrix coefficient in memory.

It is desirable to minimize the number of iterations the CG method requires to
converge to a given level of accuracy. This can be done by ``preconditioning''
the system of equations. Preconditioning amounts to multiplying on both sides
with an approximation of the inverse of the coefficient matrix and solving
this modified system. As long as the preconditioner is non-singular the
solution will be the same for the original and the preconditioned systems, but
for a well-chosen preconditioner the number of iterations can be reduced
significantly. A natural choice of the preconditioning matrix which we used to
obtain the results in this paper, is the diagonal matrix
$[\mathrm{diag}(\mathrm{A}^{\mathrm{T}}\mathrm{A})]^{-1}$, which, for a
delta-function beam, is just the inverse of the number of hits per pixel.

At every iteration we have an approximate solution $\tilde{\mathbf{x}}$ of
Eq.~(\ref{eq:ls-gauss}). We assess convergence by computing the ratio of $L_2$
norms \beq \frac{L_2[\mathbf{B}^{\mathrm{T}} \mathbf{B}\tilde{\mathbf{x}} -
\mathbf{B}^{\mathrm{T}} \mathbf{d}]} {L_2[\mathbf{B}^{\mathrm{T}} \mathbf{d}]}
\eeq where $L_2[\mathbf{x}]\equiv\sqrt{\abs{\mathbf{x}\cdot \mathbf{x}}}$.

\subsection{Standard Map-Making: Brief Review and Critique}
In order to compare our results to traditional techniques we also implemented
a traditional map-making code that solves the normal equation (Eq.~(\ref{eq:least-squares})) assuming an azimuthally symmetric beam. In this
implementation the observation matrix $\mathbf{A}$ becomes the pointing
matrix, containing only a single entry on each row corresponding to the
direction in which the main beam lobe is pointing at the time of
sampling. Standard map-making therefore
reconstructs a  map which is smoothed by an effective beam whose shape varies
as function of position on the map. This variation depends on the scanning
strategy. More precisely, at any given position on the estimated map the effective beam
shape depends on the various orientations of the beam as it passed through
this position during the scan.

For uncorrelated noise and an azimuthally symmetric beam the solution of the
normal equation is simple to compute: bin the TOD into discrete sky pixels,
summing over repeated hits, and dividing through by the number of hits per
pixel. Numerical implementations of this algorithm and its generalization to
correlated noise have been described in the literature
\cite{mapmaking}. However, all of these treatments assume azimuthally
symmetric beams. For experiments with highly asymmetric beams and where
contamination from the Galaxy is picked up in the sidelobes, we expect that
this method will not fare well against our deconvolution method which also
removes artifacts due to these optical systematics.
We use the same TOD and scan path as for our deconvolution method.  Here, the
data is binned into pixelized maps, rather than into the
$T(\Phi_2,\Theta,\Phi_1)$ grid.  Unless otherwise stated we use the HEALPix
pixelization scheme \cite{healpix} with resolution parameter $nside=64$. The
angular scale of a pixel is therefore just under $1^\circ$.  Recall that our
regularization method returns a smoothed map with an effective, azimuthally
symmetric Gaussian beam. Thus, in order to compare the two methods we must
make a similar modification to our standard map-making.  We read out the
resulting $a_{\ell m}$ of our standard map (using the HEALPix {\tt anafast}
routine), after the binning step, and modify them in the following way
\begin{equation}
a_{\ell m}' = \frac{a_{\ell m}}{B_{\ell}}G_{\ell}
\end{equation}
where $B_{\ell}$ is the beam power spectrum and $G_{\ell}$ is given in
Eq.~(\ref{eq:gauss}).

\section{Test Cases}
\label{tests}
In this section we detail our tests and comparisons of the deconvolution and
standard map-making methods.  For the purposes of testing our method, we
create several mock beam models $b_{\ell m}$. We test three possible beam
shapes which break azimuthal symmetry progressively strongly, two scanning patterns, and skies with and without Galactic emission.

The first beam is a simple model of a sidelobe; it is composed of a Gaussian
beam of $\mathrm{FWHM} = 1800'$ rotated at $90^\circ$ to another Gaussian beam
of $\mathrm{FWHM} = 180'$. Both the main beam and the sidelobe are normalized
such that they integrate to one. The second beam models a (somewhat exaggerated)  
elliptical shape, composed of two identical Gaussian beams with $\mathrm{FWHM} = 180'$
whose centers are on both sides of the optical axis, separated by $180'$. The
third beam is composed of two identical Gaussian beams ($\mathrm{FWHM}=180'$) rotated at
$140^\circ$ from each other; we refer to this as the two-beam model.  This
case is motivated by the design of the WMAP observatory \cite{BMAP03}.

We set the asymmetry parameter $m_{\rm max}$ for our three cases (sidelobe,
elliptical, and two-beam) to 8, 38, and 128, respectively.

Following \cite{WG01}, we first considered a {\it basic scan path} (BSP) in
which the beam scans the full sky on rings of constant longitude with no
rotation about its outward axis.  To be clear, for the case of the sidelobe
beam, the smaller beam follows this ringed-scan while the larger beam remains
fixed at the equatorial longitude.  Similarly, in the two-beam model, one beam
follows the ring-scan while the other rotates in smaller circles $140^\circ$
away. The central lobe therefore covers the whole sky, while the offset beam
remains within a band of $\pm 50^\circ$  
centered on the ecliptic 
plane.  The elliptical beam simply follows the ring-scan, and is oriented such
that its long axis remains perpendicular to the lines of longitude.

A more realistic observational strategy has a beam that
revisits locations on the sky in different orientations.  Therefore, we model
the one-year WMAP scan path followed by one horn.  The WMAP scan strategy also
covers 
the full sky and includes a spin modulation of $0.464$ revolutions per minute
and a spin precession of one revolution per hour \cite{BMAP03}.  We used a
scaled-down model of the WMAP scan in which the spin modulation is $0.00232$
revolutions per minute and a step size of about 46 seconds (roughly 562
samples per period).  This produces a pattern very similar to the
spirograph-type pattern shown in Fig.~4 of \cite{BMAP03}.  We refer to this as
the WMAP-like scan path (WSP).  The WSP has about six times as many samples as
the BSP. %For comparison, %the basic scan strategy takes 513 %samples per ring
For this strategy, the spirograph pattern is followed by the small beam of the
sidelobe, the elliptical beam, and {\em both} beams of the two-beam model. In
the two-beam case both beams are  offset from the spin axis of the satellite,
to mimic the WMAP scanning geometry. It is not differential in nature, since both beams 
have positive weight.

We test each beam (sidelobe, elliptical, and two-beam) with both scanning
patterns (BSP and WSP) on a sky without Galactic emission.  We refer to these
as the six main test cases.

In reality, CMB experiments will also pick up signal from the Galaxy.  We use
the first-year WMAP Ka-band temperature map, degraded to an $nside$ of 64 and
smoothed with a Gaussian beam of $\mathrm{FWHM}=180'$ as our model of the true
sky with Galactic emission.  For our last test case, we convolve it with the
sidelobe beam.

For each test case, we assume that the beamshapes of the instrument are known
and use the deconvolution method to deconvolve the map with the same beam that
the true sky was originally convolved with.  We attempt to recover features in
the map corresponding to the smallest scale features of our test beams. We
therefore 
set the width of our regularization kernel, represented by the matrix
$\mathbf{G}$ in Eqs.~(\ref{eq:gauss}) and (\ref{eq:ls-gauss1}), to
$\mathrm{FWHM}=180'$ in every case. We compare our map estimates to the true
sky, smoothed by the regularization kernel.  When we refer to the ``true'' sky
in the following we mean this kernel-smoothed input sky.

\section{Results and Discussion}
\label{results}
We present the results of the deconvolution algorithm for the six main test
cases in the form of residual maps.  We compare these results to the results
from standard map-making by examining their power spectra and by calculating
the root mean square (RMS) difference between the estimated and true sky. For
the tests that include the Galactic signal we
show the actual map estimates.

\begin{figure*}
\includegraphics[width=.6\textwidth, keepaspectratio,
  angle=90,origin=lB]{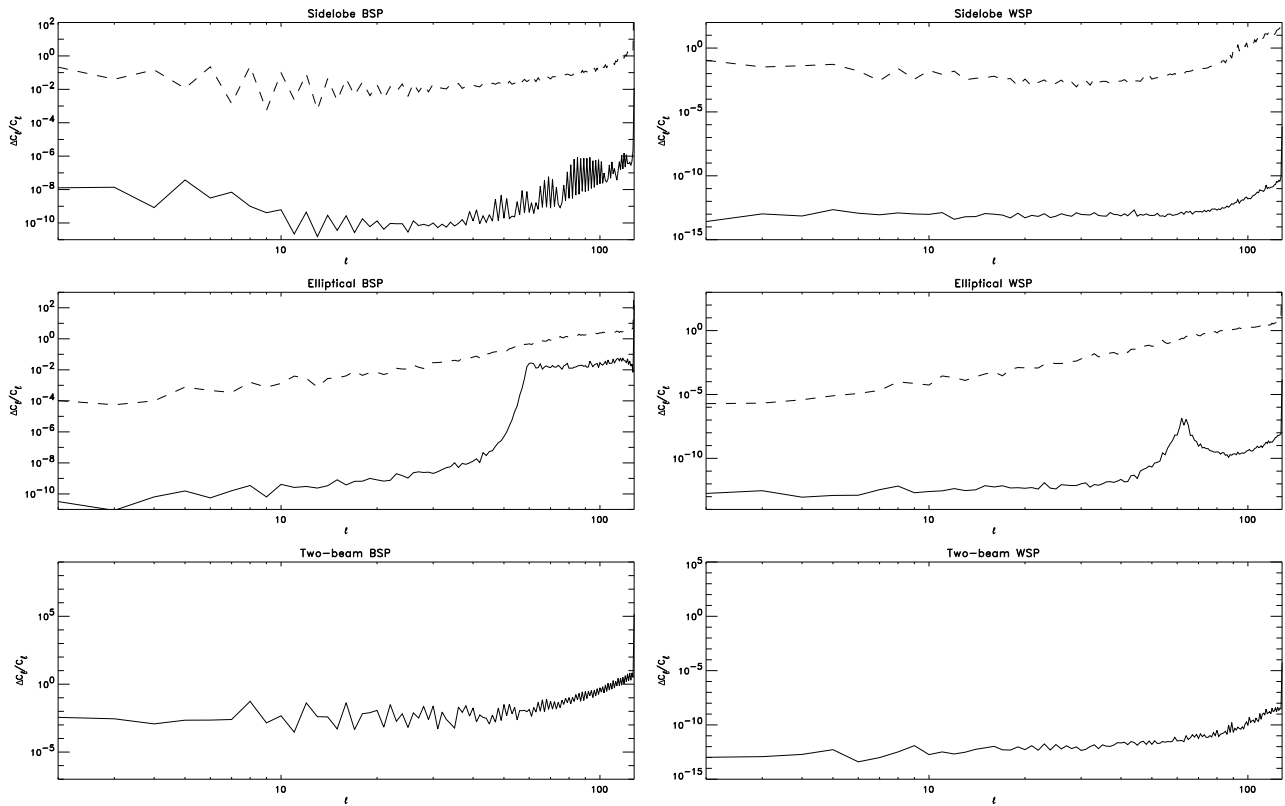}
\caption{Ratios of the spectra of the residual map to the spectrum of the input
  map for each of the beam models and both scanning strategies.  The BSP
  results are plotted in the left column and the WSP results are plotted in
  the right column.  Results for the sidelobe, elliptical and two-beam beam
  are shown in the top, middle, and bottom panels, respectively.  The solid
  lines correspond to deconvolved spectra and the dashed lines correspond to
  the standard spectra.}
\label{fig:cls}
\end{figure*}

In Fig.\ \ref{fig:cls} we plot ratios of the power spectra of the residual
maps (both standard and deconvolved) and the power spectrum of the input map.
The BSP (WSP) results are plotted in the left (right) column. The solid
(dashed) lines represent the relative difference in $C_{\ell}$ between the
deconvolved (standard) map and true sky map. The standard map-making algorithm
failed to give meaningful results for the two-beam test. We therefore excluded
this case from the plot.

For  all cases we chose to present the
results  after
a fixed number of iterations to show the impact of scanning strategy and
beam pattern on the condition number of the map-making equations. We find that the deconvolution algorithm outperforms standard map-making by
orders of magnitude in accuracy.

For a  fixed number of iterations, the BSP tests performed less
well than the WSP tests. The two-beam BSP
and, to a lesser extent, the elliptical beam BSP test cases have not
converged to sufficient  accuracy. 

There are several possible causes for this behaviour. The BSP leads to 
an  extremely 
non-uniform sky coverage. Also, the BSP visits each pixel in a
narrow range of beam orientations. Further, the number of sky samples is
smaller for the BSP case than for the WSP case (as noted in section
\ref{tests}). All of these aspects can contribute to increasing the condition 
number of the normal equation, which in turn leads to smaller error decay per
iteration of the preconditioned CG solver.

\begin{figure*} 
\includegraphics[ width=.3\textwidth, keepaspectratio,
  angle=90,origin=1B]{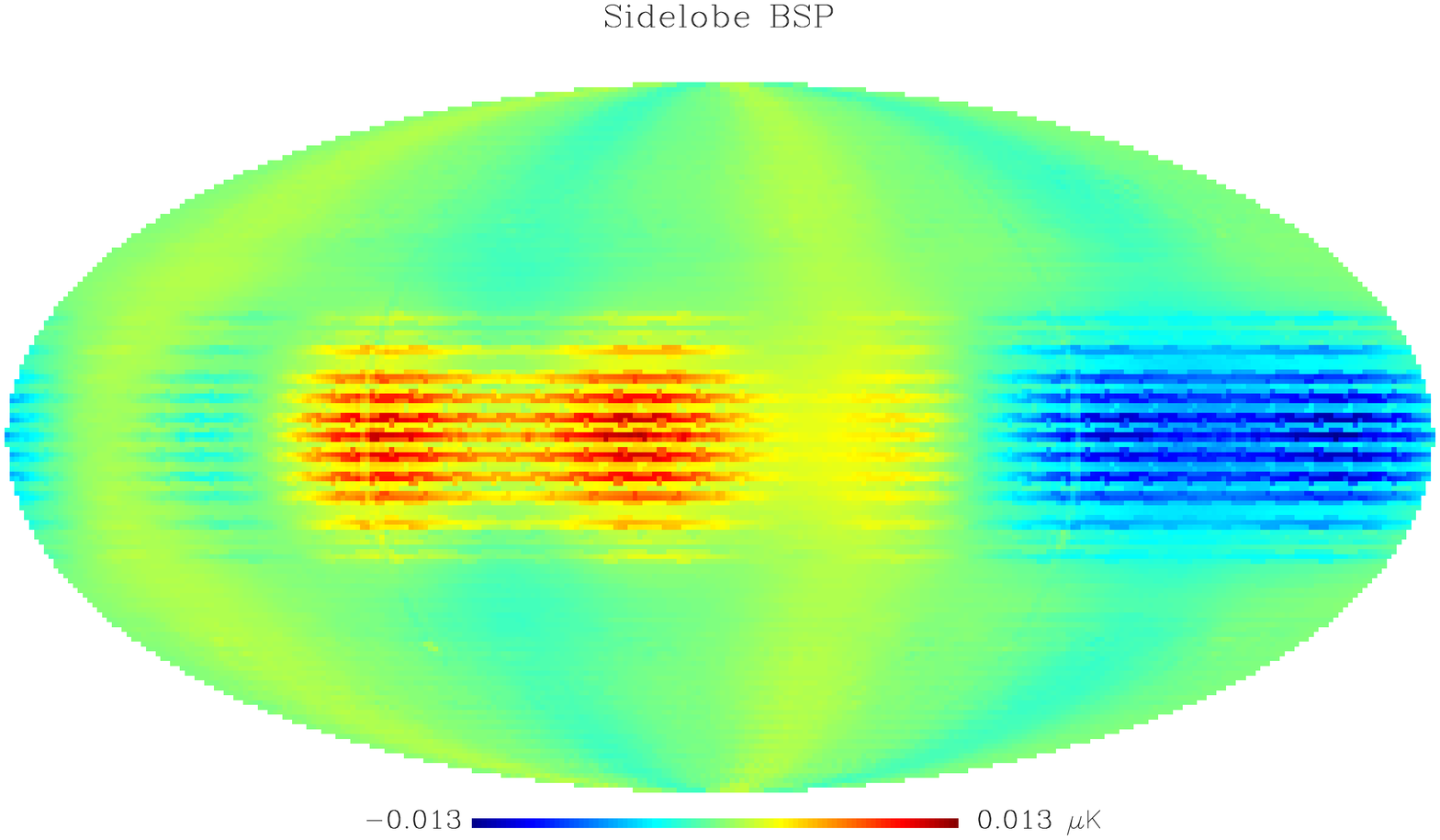}
  \includegraphics[ width=.3\textwidth, keepaspectratio,
  angle=90,origin=1B]{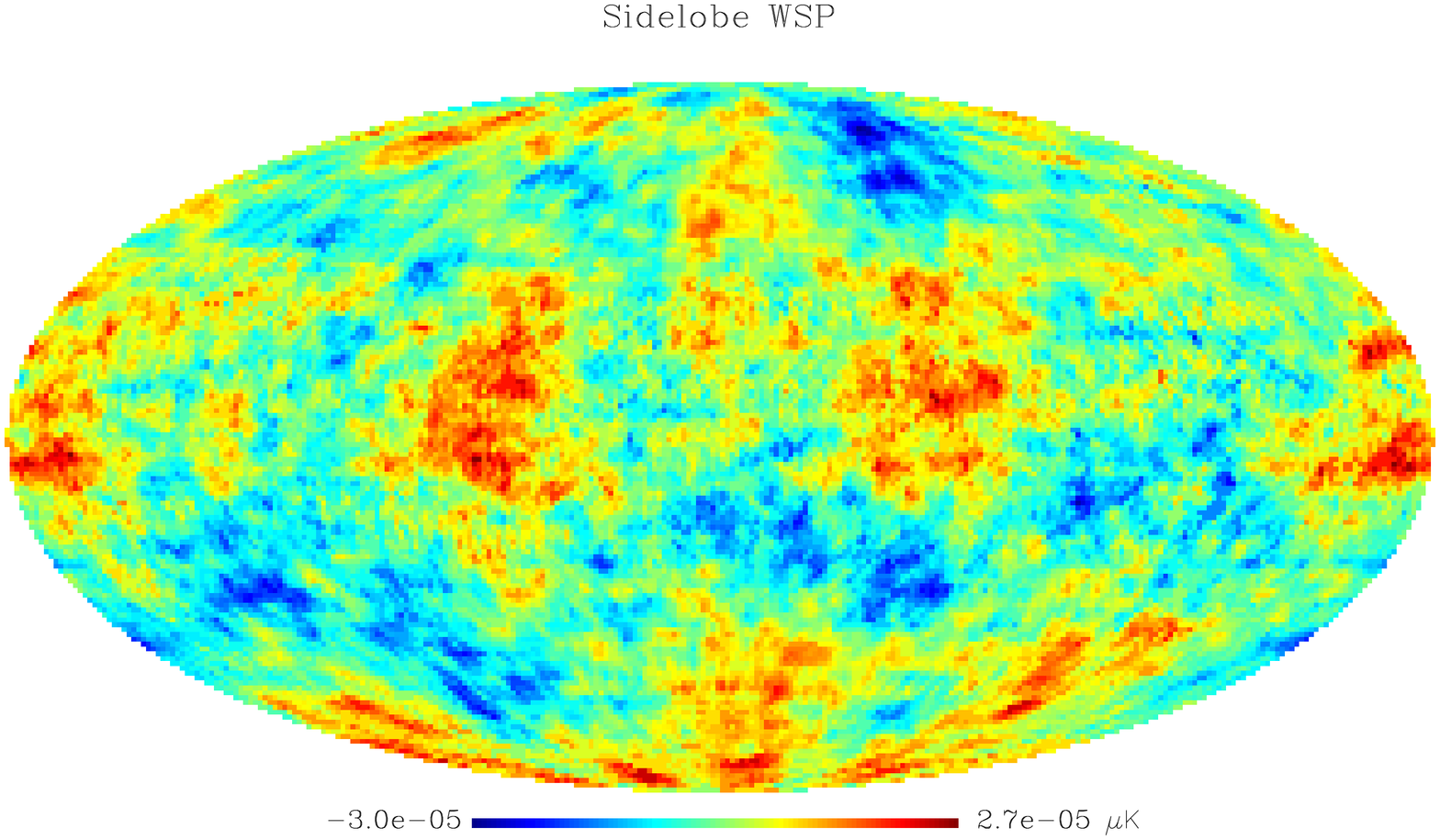}
  \includegraphics[ width=.3\textwidth, keepaspectratio,
  angle=90,origin=1B]{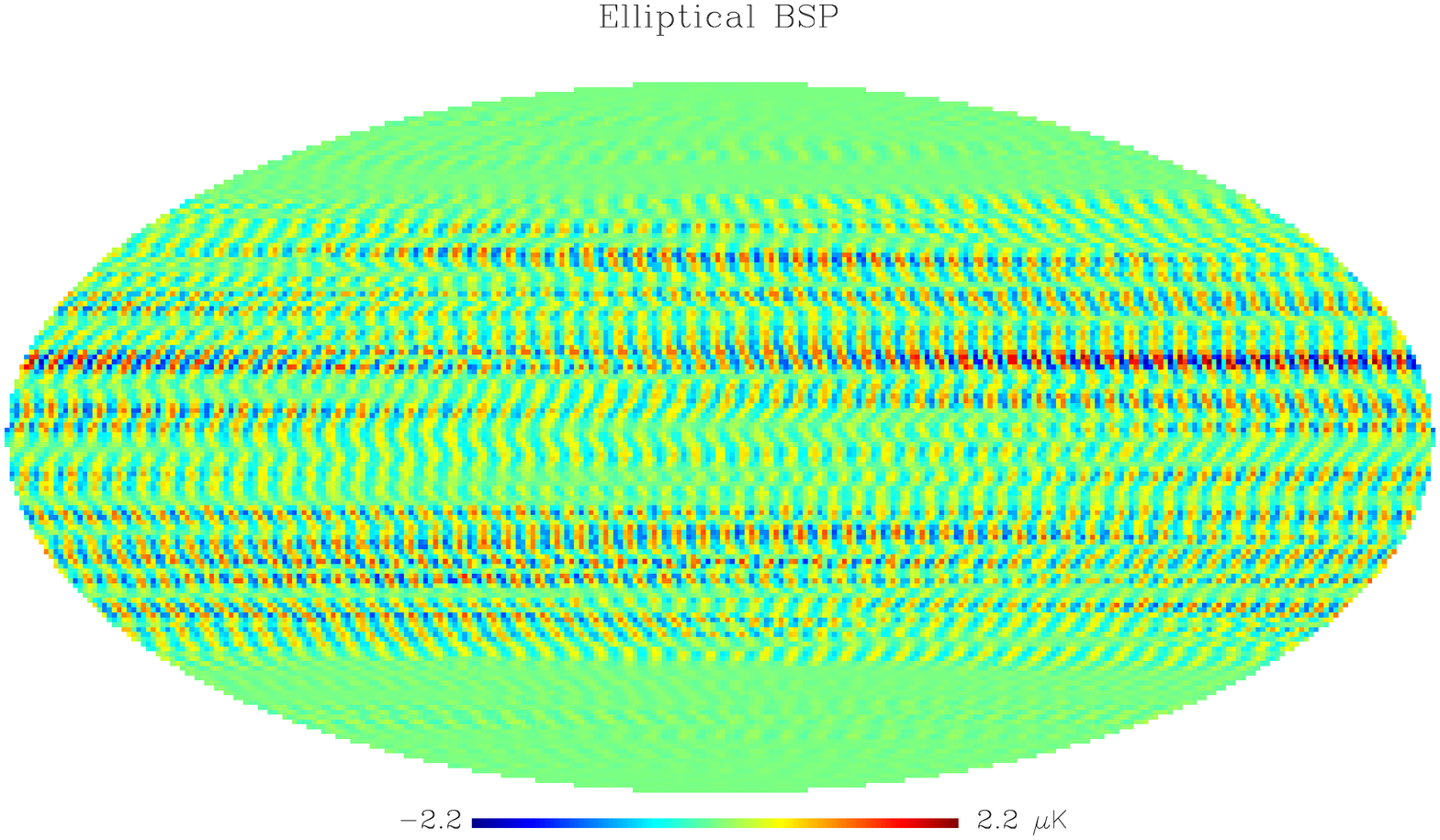}
  \includegraphics[ width=.3\textwidth, keepaspectratio,
  angle=90,origin=1B]{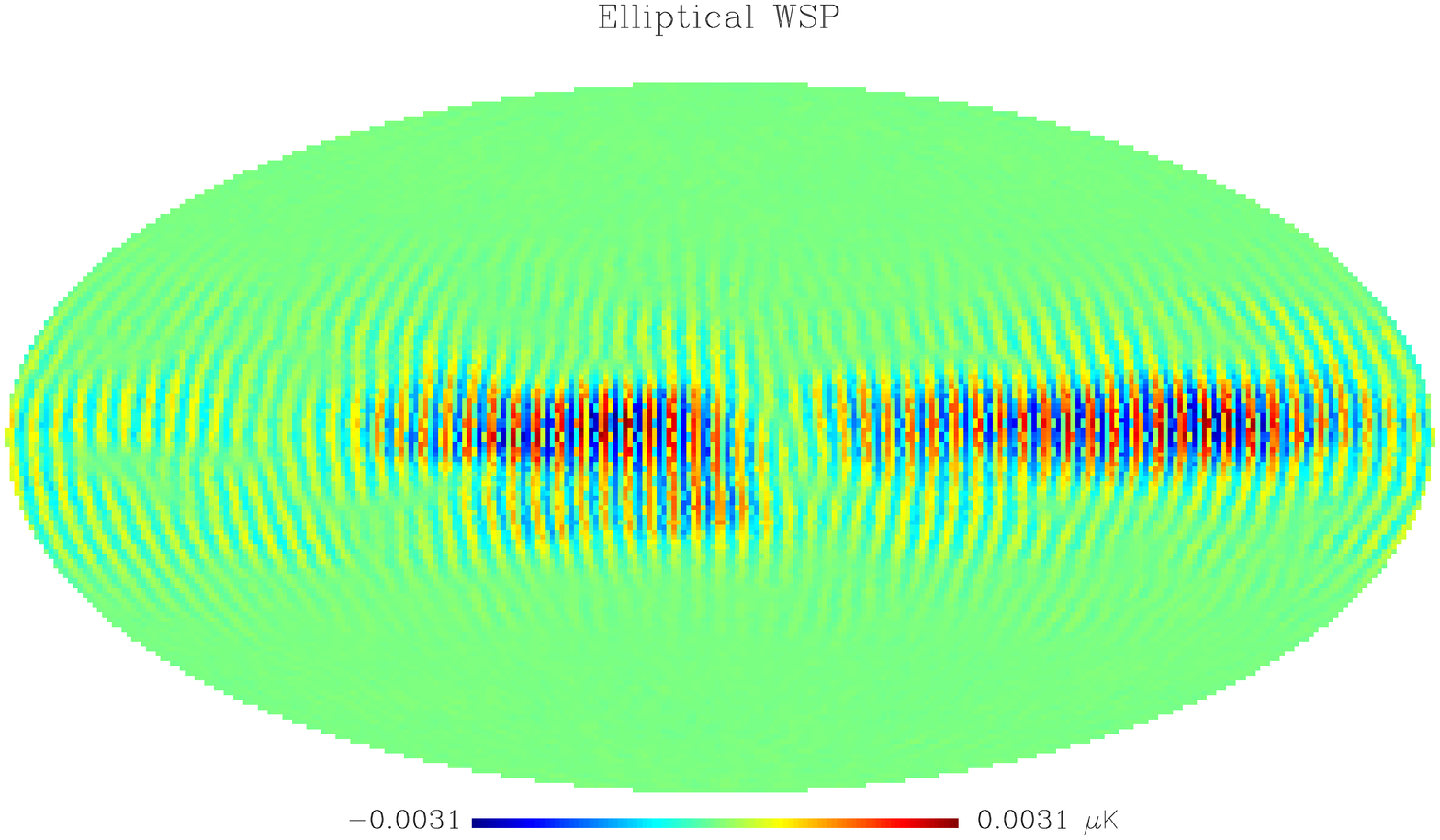}
  \includegraphics[ width=.3\textwidth, keepaspectratio,
  angle=90,origin=1B]{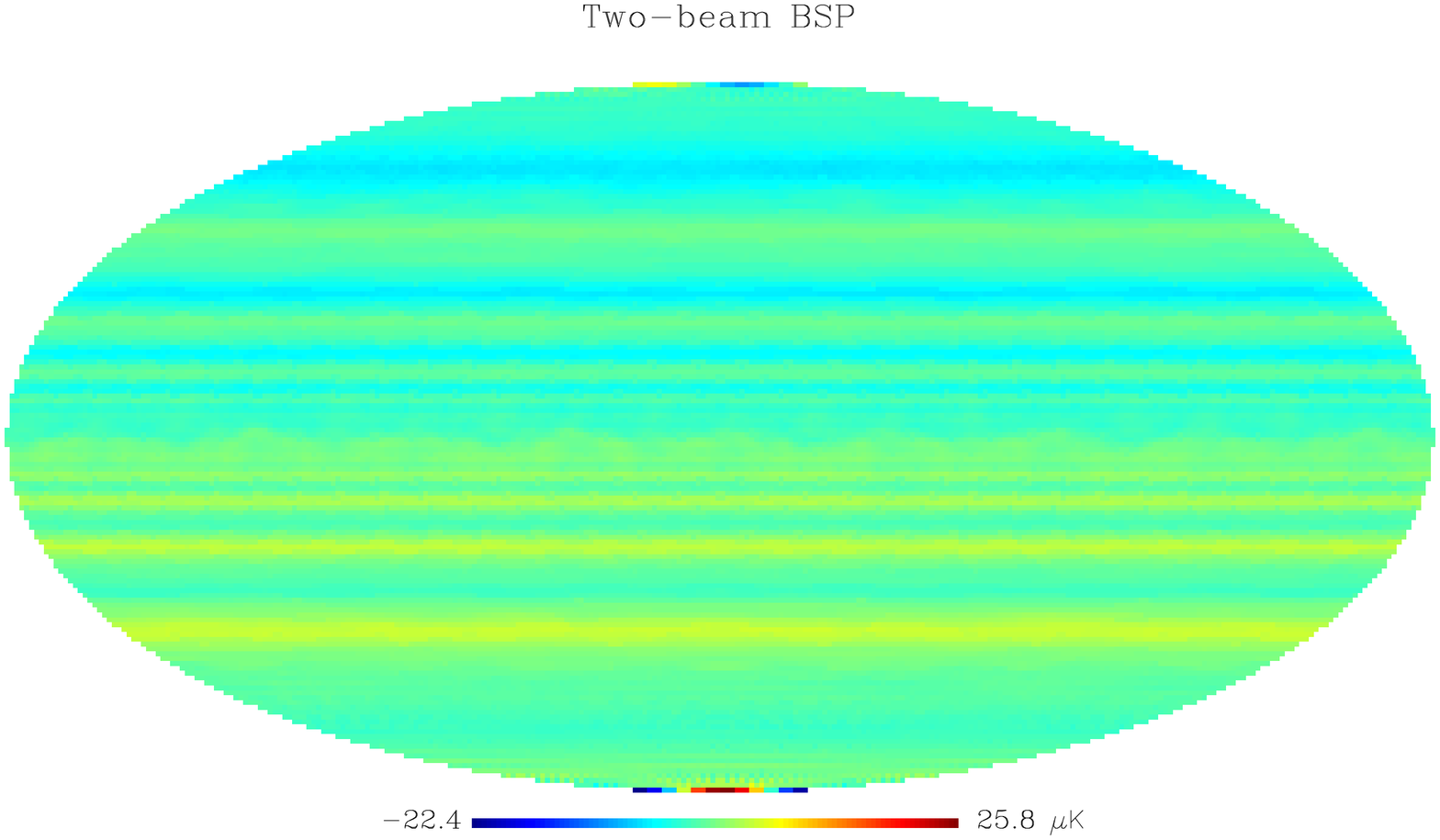}
  \includegraphics[ width=.3\textwidth, keepaspectratio,
  angle=90,origin=1B]{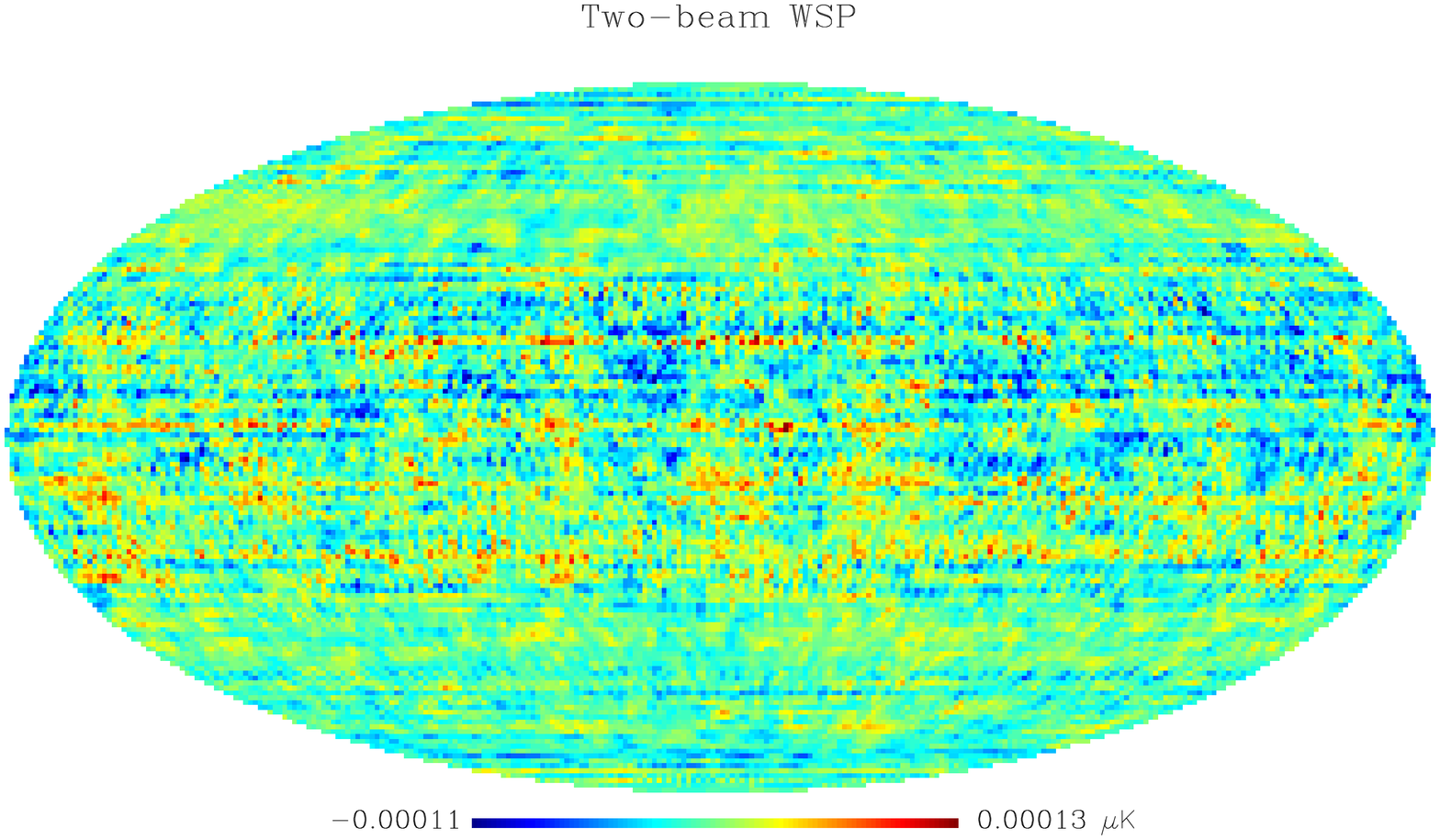}
  \includegraphics[ width=.3\textwidth, keepaspectratio,
  angle=90,origin=1B]{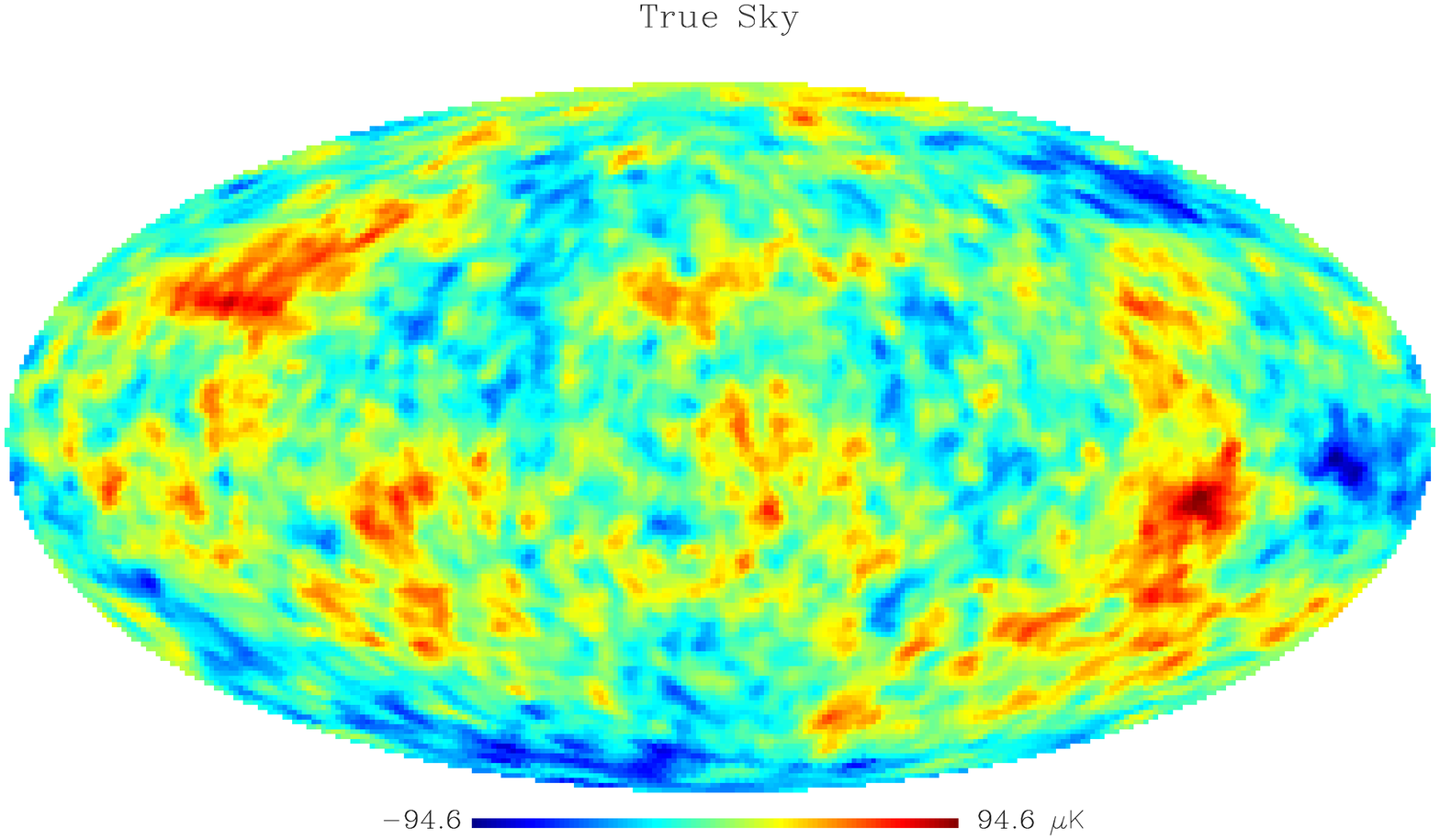}
\caption{Residuals after the first 100 iterations are shown in the first
  three rows.  The figures on the left are for the basic scan path the ones on
  the right for the WMAP-like scan path.  First, second, and third rows
  correspond to the sidelobe, elliptical and two-beam beams, respectively.
  The true sky is shown in the fourth row. Note that solutions of the two-beam
   BSP and, to a lesser extent, the elliptical beam BSP test cases have not
  converged to sufficient  accuracy. We chose to present the results
  for  all cases after
  a fixed number of iterations to show the impact of scanning strategy and
  beam pattern on the condition number of the map-making equations.}
\label{fig:resmaps}
\end{figure*}

In Table 1 we summarize the RMS difference between the reconstructed and true
sky. The RMS values are computed using the standard deviations the residual
and true maps:
\begin{equation}
\mathrm{RMS} = \frac{\mbox{stdev(residual map)}}{\mbox{stdev(true map)}}
\end{equation}
where 
\begin{equation}
%\mbox{residual map} = \frac{\mbox{true map}}{\mbox{stdev(true map)}} -
%\frac{\mbox{estimated map}}{\mbox{stdev(estimated map)}}.
\mbox{residual map} = {\mbox{estimated map} - \mbox{true map}}.
\end{equation}
The RMS values reflect the trends seen in the spectra in
Fig.~\ref{fig:cls}.  The residual maps are shown in Fig.~\ref{fig:resmaps}.
In order that the scale of the axes on the residual maps are meaningful, we
also show the true sky map.

\begin{table}[h]
%\centering
\begin{tabular}{|l||l|l||l|l|}
\hline &\multicolumn{2}{l|}{BSP}&\multicolumn{2}{l|}{WSP}\\ \cline{2-5} Beam
 &Standard&Deconvolved&Standard&Deconvolved\\ \hline\hline 
Sidelobe& 0.257467 & 0.000111903 & 0.178828 & 3.13741e-07 \\
Elliptical& 0.186262 & 0.0207830 & 0.129715 & 2.25602e-05 \\
Two-beam& N/A & 0.102778 & N/A & 1.08579e-06 \\ \hline
\end{tabular}
\label{tab:rms}
\caption{Fractional RMS error for each of the six main test cases.}
\end{table}

\begin{figure*}
\includegraphics[ width=.9\textwidth, keepaspectratio,
  angle=0,origin=1B]{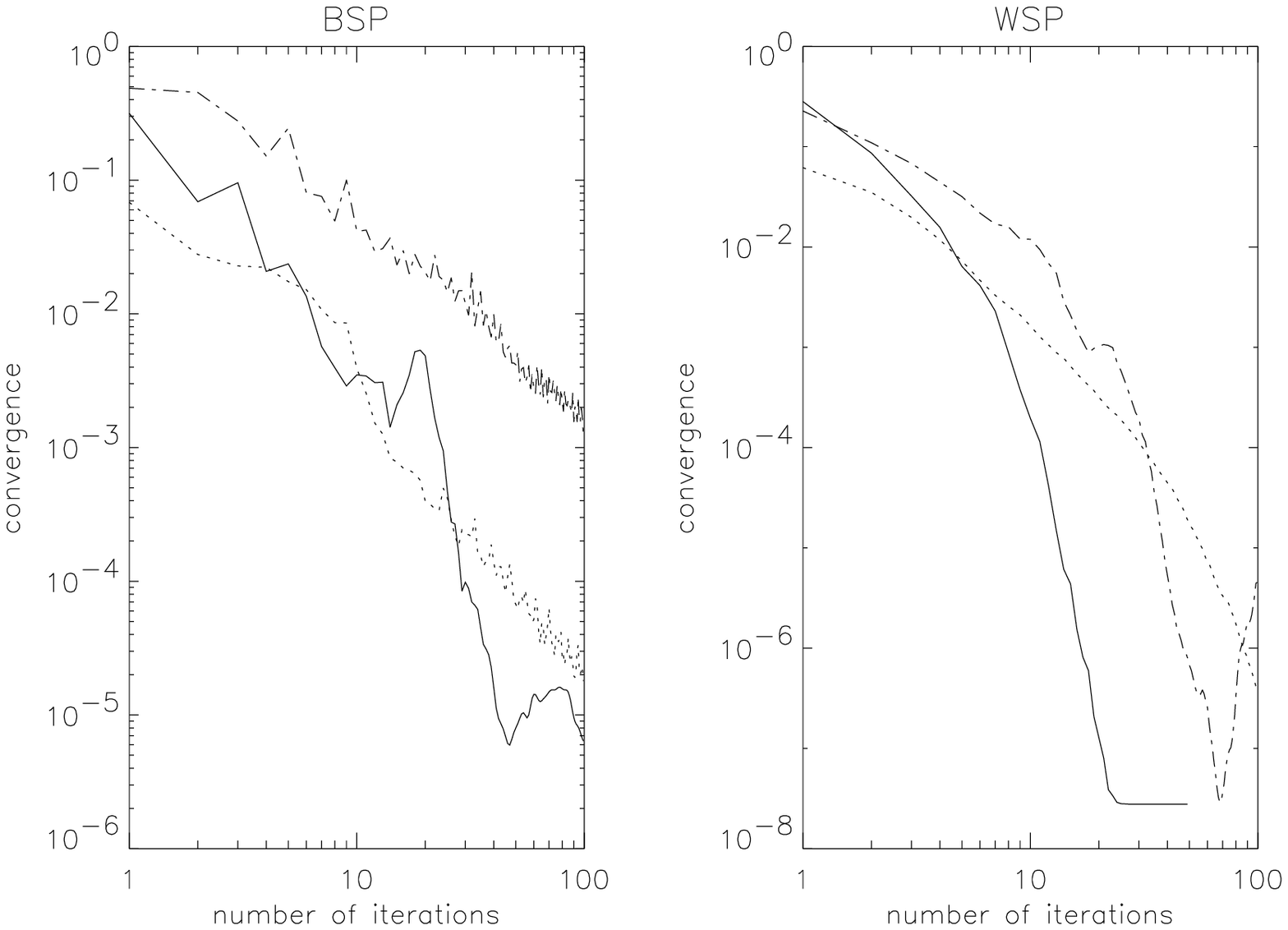}
\caption{Convergence rates of the preconditioned conjugate gradient solver for
  each test case.  The left panel refers to the basic 
  scan path and the right panel to the WMAP-like scan path.  The solid lines
  correspond to the sidelobe beam, dotted lines to the elliptical beam, and
  dot-dashed to the two-beam model.}

\label{fig:converge}
\end{figure*}

 Achieving a stably converging iterative solution method for the deconvolution
 problem is 
a success of our regularization technique. The convergence of our iterative solver as function of iteration number is
plotted in Fig.~\ref{fig:converge}. 

 In order to be able to compare the
performance of our method for different beam patterns and scanning strategies
we make the deliberate 
choice of limiting the number of iterations to 100 and comparing the best
results obtained up to this point. Since our error estimate continues to drop
stably 
(except in two cases, where we reach the single precision numerical accuracy
floor after $\sim 25$ and $\sim 70$ iterations) it is clear that the accuracy
of the reconstruction can be improved by allowing the system to iterate further, or by
choosing a more sophisticated pre-conditioner.

\begin{figure*}
\includegraphics[ width=.4\textwidth, keepaspectratio,
  angle=90,origin=1B]{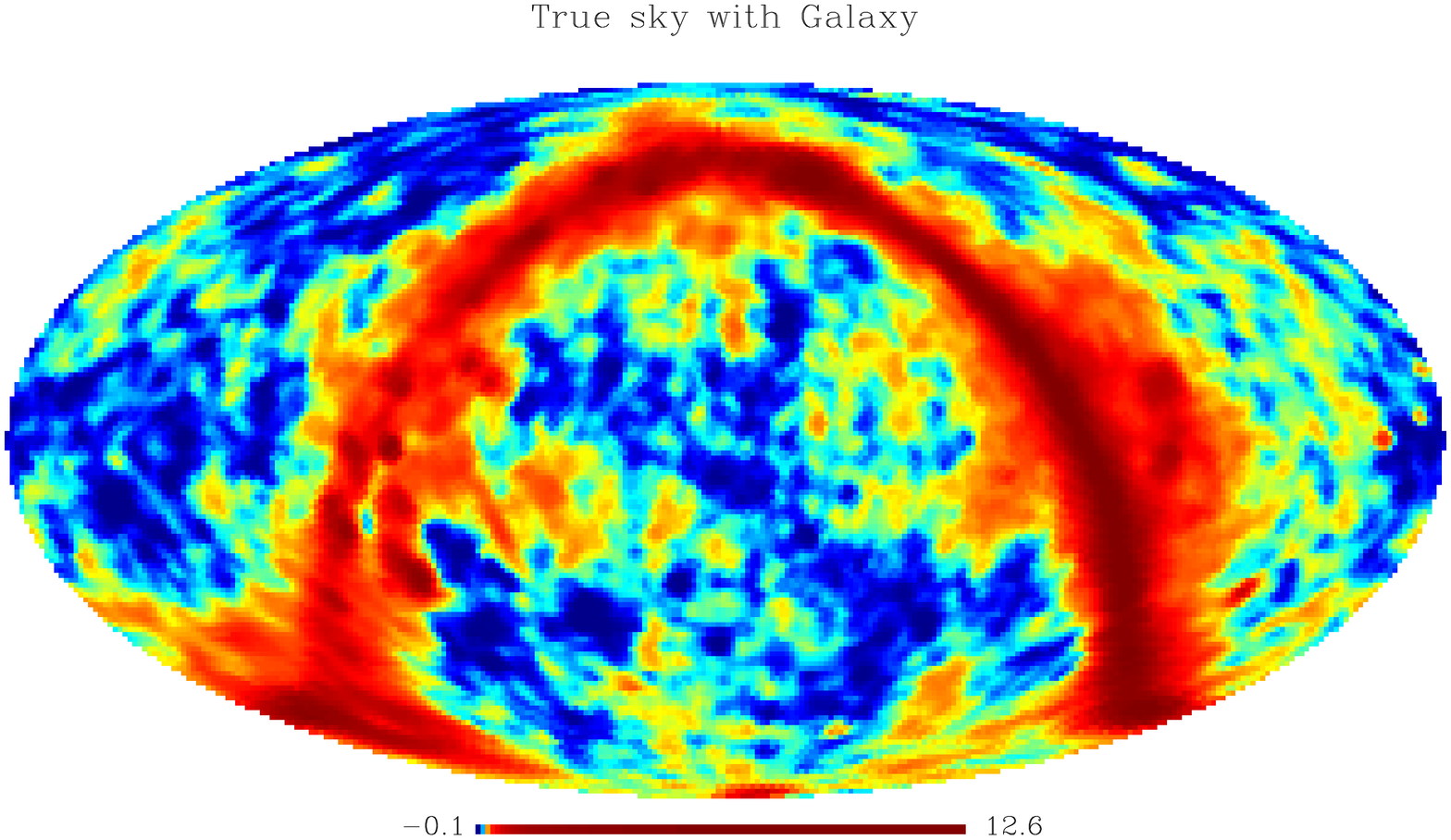}
  \includegraphics[ width=.4\textwidth, keepaspectratio,
  angle=90,origin=1B]{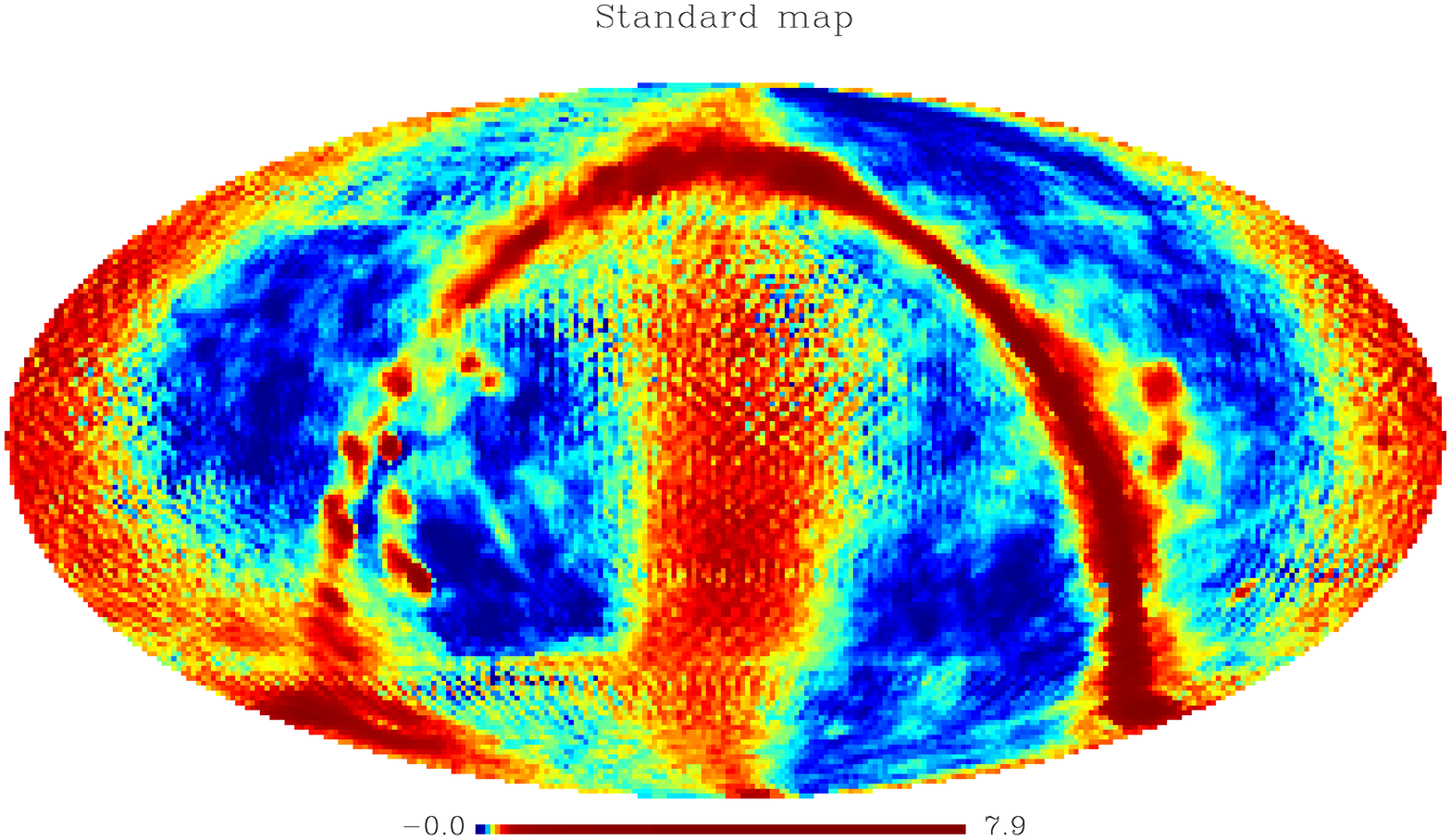}
  \includegraphics[ width=.4\textwidth, keepaspectratio,
  angle=90,origin=1B]{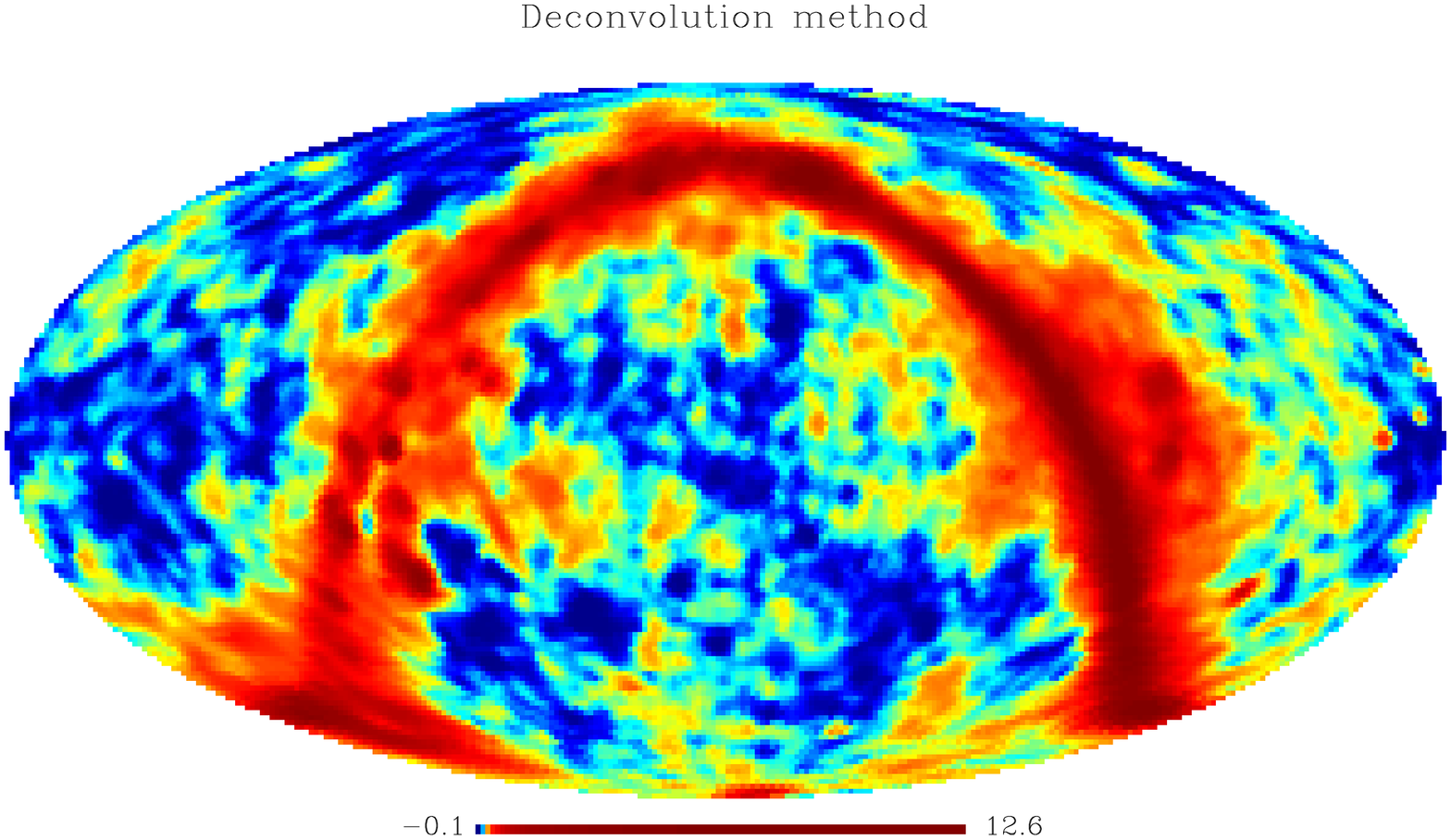}
\caption{Deconvolving the effects of a large sidelobe in simulated
  observations of the WMAP Ka band map, using the coarsened WMAP scanning
  strategy described in the text. The top map is the input sky map, the middle map is the standard 
map-making result, and the bottom map is the deconvolved result.}
\label{fig:gal}
\end{figure*}

Our final test case consists of a model sky with Galaxy emission convolved
with the sidelobe beam over the WSP.  We present the output maps of both the
standard and deconvolution methods in Fig.~\ref{fig:gal}, where the maps are
shown in ecliptic coordinates.  One can see that the standard map contains a
distorted image of the Galaxy and that the deconvolved map is virtually
identical to the true map.

\section{Conclusions and Future Work}
\label{conclusions}
We have presented a deconvolution map-making method for data from scanning CMB
telescopes. Our methods remove artifacts due to beam asymmetries and far
sidelobes.  We compare our technique with the standard map-making method and
demonstrate that the true sky is recovered with greatly enhanced accuracy via
the 
deconvolution method. Deconvolution map-making recovers features of the CMB
sky on the smallest scale of the beam, thereby achieving a form of
super-resolution imaging. This extracts more of the information content in CMB
data sets.

One of the key difficulties encountered in deconvolution problems is that the systems of
linear equations we need to solve are very nearly singular. We solve this
problem by introducing  a regularization method 
which allows us to solve the systems stably and recover maps
at a uniform resolution and with an effective beam that is azimuthally
symmetric and has a Gaussian profile. 

We tested the convergence speed of two particular scanning strategies and
found that the 
WMAP-like scan is superior to the basic scan in both rate of convergence and
true-sky recovery.  We hypothesize that this is due to the nature of the BSP,
where the poles are the location of the only beam crossings and receive many
more hits than the rest of the sky. In addition, our implementation of the BSP
had a smaller number of samples overall than our implementation of the
WSP.

We have also shown the relevance of this algorithm to the WMAP mission by
demonstrating its operation using a WMAP-like scanning strategy and a two-beam
model which, while not differential, resembles the telescope orientations of
the WMAP spacecraft. Our results underline the qualities of the WMAP scanning
strategy compared to a BSP strategy for deconvolution map-making.

In order to decouple from issues that are not directly related to the optical
performance of CMB instruments we did not consider the effects of noise in our
simulations. For a realistic assessment of the performance of our methods on
real data this needs to be added. In particular, the choice of scale for the
regularization kernel will depend on weighing the benefits of increased
resolution against increased high-frequency.

Recently several groups have published CMB polarization results on the EE
power spectrum \cite{polarization}. WMAP released an all-sky analysis of the
TE cross-correlation in the first year data \cite{WMAPpol}.  We eagerly
anticipate the large-angle polarization data from WMAP in the impending
release of the second year data. In a few years' time, polarimeters on board
the Planck satellite will collect data. Owing to the difficulty of
separating the two polarization modes, we expect that polarimetry experiments will be
very sensitive to beam asymmetries and stray light.  Future measurements of
the tiny B-mode polarization will require both exquisite instruments and
sophisticated analysis tools.  We hope that the generalization of our methods
to polarized map-making will be useful for making maps of the polarized
microwave sky.  It has already been shown \cite{C00} that little modification
to the Wandelt-G\'{o}rski method of fast all-sky convolution is needed to
accomodate polarization data.

\acknowledgments
We thank the performance engineering group at NCSA, in particular Gregory
Bauer, for stimulating conversations and help with optimizing our code. BDW
gratefully acknowledges 
funding from 
the National Center for Supercomputing Applications and the Center for
Advanced Studies. This work was partially supported by
NASA contract JPL1236748, by the
National Computational Science Alliance under AST300029N and the University of
Illinois.   We utilized the IBM pSeries 690 Cluster
\texttt{copper.ncsa.uiuc.edu}.

 \end{document}

%% file: defs.tex
\newcommand\ba{\begin{array}}
\newcommand\ea{\end{array}}
\newcommand\bc{\begin{center}}
\newcommand\ec{\end{center}}
\newcommand\be{\begin{enumerate}}  
\newcommand\ee{\end{enumerate}}  
\newcommand\bi{\begin{itemize}}  
\newcommand\ei{\end{itemize}}  
\newcommand\bd{\begin{description}}  
\newcommand\ed{\end{description}}  
\newcommand\beq{\begin{equation}}  
\newcommand\eeq{\end{equation}}  
\newcommand\beqa{\begin{eqnarray}}  
\newcommand\eeqa{\end{eqnarray}}

\newcommand\eg{{\em e.g.\ }}

\newcommand\mathC{\mkern1mu\raise2.2pt\hbox{$\scriptscriptstyle|$}
                {\mkern-7mu\rm C}}

\newcommand\abs[1]{\left\vert {#1}\right\vert}

\def\exl{\raise1pt\hbox{$\scriptstyle<$}}
\def\exr{\raise1pt\hbox{$\,\scriptstyle>$}}

\newcommand\order[1] {{{\cal O}\! \left( #1 \right)} }

\newcounter{theo}[section]
\renewcommand{\thetheo}{\thesection.\Roman{theo}}

\newcounter{cor}
\renewcommand{\thecor}{\thetheo.\roman{cor}}

\newcounter{exa}

\newcounter{def}[section]
\renewcommand{\thedef}{\thesection.\arabic{def}}

\newcounter{rem}